\documentclass[usenatbib]{mnras}

\usepackage{graphicx,times, amsmath, epsfig, color}
\usepackage[export]{adjustbox}

\newcommand{\cmfast}{\textsc{\small 21CMFAST}}

\newcommand{\Ts}{T_{\rm S}}

\newcommand{\nf}{x_{\rm HI}}

\newcommand{\Msun}{M_\odot}

\newcommand{\Tvirmin}{T_{\rm vir}^{\rm min}}

\newcommand{\Tcmb}{T_\gamma}

\newcommand{\delT}{\delta T_b}

\newcommand{\delNL}{\delta_{\rm nl}}

\newcommand\lsim{\mathrel{\rlap{\lower4pt\hbox{\hskip1pt$\sim$}}
        \raise1pt\hbox{$<$}}}
\newcommand\gsim{\mathrel{\rlap{\lower4pt\hbox{\hskip1pt$\sim$}}
        \raise1pt\hbox{$>$}}}
\def\myputfigure#1#2#3#4#5%
{\vskip#5pt\makebox[0pt]{\hskip#2in
\includegraphics[width=#3\textwidth]{#1}}\vskip#4pt\hfill}

\pdfoutput=1

\begin{document}

\title[21-cm signatures of X-ray absorption in the first galaxies]{High Mass X-ray Binaries and the Cosmic 21-cm Signal:\\
 Impact of Host Galaxy Absorption}

\author[Das et al.]{Arpan Das$^1$,
 Andrei Mesinger$^{1}$\thanks{andrei.mesinger@sns.it},
 Andrea Pallottini$^{1,2,3}$,
 Andrea Ferrara$^{1}$,
 John H. Wise$^{4}$\\
 $^{1}$Scuola Normale Superiore, Piazza dei Cavalieri 7, 56126 Pisa, PI, Italy\\
 $^{2}$Cavendish Laboratory, University of Cambridge, 19 J. J. Thomson Ave., Cambridge CB3 0HE, UK\\
 $^{3}$Kavli Institute for Cosmology, University of Cambridge, Madingley Road, Cambridge CB3 0HA, UK\\
 $^{4}$Center for Relativistic Astrophysics, Georgia Institute of Technology, 837 State Street, Atlanta, GA 30332, USA
}

\voffset-.6in

\maketitle

\begin{abstract}
By heating the intergalactic medium (IGM) before reionization, X-rays are expected to play a prominent role in the early Universe. The cosmic 21-cm signal from this ``Epoch of Heating'' (EoH) could serve as a clean probe of high-energy processes inside the first galaxies.
Here we improve on prior estimates of this signal by using high-resolution hydrodynamic simulations to calculate the X-ray absorption due to the interstellar medium (ISM) of the host galaxy, typically residing in halos with mass $10^{7.5-8.5} \Msun$ at $z\sim$ 8--15.  X-rays absorbed inside the host galaxy are unable to escape into the IGM and contribute to the EoH.
We find that the X-ray opacity through these galaxies can be approximated by a metal-free ISM with a typical column density of $\log[N_{\rm HI}/{\rm cm^{-2}}] = 21.4^{+0.40}_{-0.65}$., with the quoted limits enclosing 68\% of the sightline-to-sightline scatter in the opacity.
We compute the resulting 21-cm signal by combining these ISM opacities with public spectra of high-mass X-ray binaries (thought to be important X-ray sources in the early Universe).
Our results support ``standard scenarios'' in which the X-ray heating of the IGM is inhomogeneous, and occurs before the bulk of reionization. The large-scale ($k\sim0.1$ Mpc$^{-1}$) 21-cm power reaches a peak of $\approx 100$ mK$^2$ at $z\sim$ 10--15, with the redshift depending on the cosmic star formation history.
This is in contrast to some recent work, motivated by the much larger X-ray absorption towards local HMXBs inside the Milky Way.
Our main results 
can be reproduced by approximating the X-ray emission from HMXBs with a power-law spectrum with energy index $\alpha\approx1$, truncated at energies below 0.5 keV.
\end{abstract}


\section{Introduction}\label{sec:intro}

The cosmic 21-cm signal promises to be a remarkable probe of the early Universe. Since the signal is sensitive to the thermal and ionization state of the intergalactic medium (IGM), upcoming low frequency interferometers will allow us to study the unseen population of galaxies and black holes during the first billion years of our Universe, prior to the completion of cosmic reionization. First generation interferometers, such as the Low Frequency Array (LOFAR; \citealt{vanHaarlem13})\footnote{\url{http://www.lofar.org/}}, the Murchison Wide Field Array (MWA; \citealt{Tingay13})\footnote{\url{http://www.mwatelescope.org/}} and the Precision Array for Probing the Epoch of Reionization (PAPER; \citealt{Parsons10})\footnote{\url{http://eor.berkeley.edu/}} are currently taking data, hoping for a statistical detection of the epoch of reionization (EoR). Second generation instruments, the Hydrogen Epoch of Reionization Array (HERA; \citealt{DeBoer16})\footnote{\url{http://reionization.org}}, and eventually the Square Kilometre Array (SKA; \citealt{Koopmans15})\footnote{\url{https://www.skatelescope.org}} will have the sensitivity and frequency coverage to probe the 21-cm signal out to the birth of the first galaxies at $z\sim30$.

Since the IGM is seen in contrast against the CMB, the detectability of the signal is strongly dependent on their relative temperatures. The 21-cm line is seen in absorption while the IGM is colder than the CMB; when the IGM is heated to temperatures above the CMB, the signal transitions to being seen in emission. Simple estimates based on the observed X-ray luminosity ($L_X$) to star-formation rate (SFR) in galaxies out to $z\lsim4$ (e.g. \citealt{MGS12_HMXB, Basu-Zych13}), suggest that X-rays from early star-forming galaxies were sufficient to heat the IGM to temperatures above the CMB prior to reionization (e.g. \citealt{Furlanetto06, MO12}). The fact that this ``Epoch of Heating'' (EoH) likely precedes the EoR, simplifies the interpretation of the 21-cm signal. The pre-EoR signal, while the IGM is still seen in absorption against the CMB, could thus serve as a clean probe of the X-ray processes inside the first galaxies.

The dominant source of these early X-rays is usually assumed to be high-mass X-ray binaries (HMXBs), which should appear only a few Myr after the first stars (e.g. \citealt{Fragos12}). Alternatively, the hot interstellar medium (ISM), heated by supernovae (SNe) and cooling through Bremsstrahlung and metal line cooling, can play a prominent role (e.g. \citealt{Pacucci14}), given that the hot ISM and HMXBs are observed to have comparable soft band\footnote{Throughout, ``soft band'' will refer to X-ray energies below $\sim2$ keV, motivated by the detectors on the {\it Chandra} X-ray telescope. Note that the mean free path through the neutral IGM can be expressed as (e.g. \citealt{FOB06, McQuinn12}): $\lambda_{\rm X} \sim 40 ~ ( E_{\rm X}/{\rm 0.5 ~ keV})^{3} ( (1+z)/15)^{-2} ~ {\rm comoving ~ Mpc}$. Equating this to the Hubble length and solving for the photon energy, one obtains that only photons with energies less than $E_{\rm hubble} \sim 2 ( (1+z)/15 )^{1/2} {\rm keV}$ are likely to interact with the IGM. Therefore the EoH is only sensitive to the rest-frame SED in the soft band.} luminosities in local star-forming galaxies (e.g. \citealt{Strickland00, Grimes05, Owen09, MGS12_ISM, LW13}).

Additionally, \citet{Xu14} postulated that the top-heavy initial mass function expected for the first, metal-free (PopIII) stars might have boosted the X-ray production in excess of estimates based on local IMFs. Although PopIII stars likely only dominate the cosmic SFR in the very first stages of the Cosmic Dawn (e.g. \citealt{SFD11,Pallottini14, KW16}), their X-ray luminosity per unit star formation
could be large if they have a high binary fraction and/or strong accretion onto the majority of their remnants (e.g. \citealt{TAO09, SGB10, SB13}). Larger black holes at high-$z$, if surrounded by enough gas to fuel accretion (e.g. so-called mini-QSOs) could also contribute to the EoH (e.g. \citealt{Yue13, GCD16}). An even more exotic source of heating is through the annihilation of dark matter particles, which can dominate the IGM heating rates in certain particle models (e.g. \citealt{EMF14, Honorez16}).

The bolometric X-ray luminosity per unit star formation of the first galaxies can be calibrated to observations of local galaxies or modeled with stellar population synthesis. However, the 21-cm signal during the EoH depends also on the X-ray spectral energy distribution (SED). Since the photo-ionization cross-sections of hydrogen and helium are strong functions of photon energy, low energy X-rays are able to more efficiently heat the IGM, provided they manage to escape the host galaxy. The SED also impacts the inhomogeneity of the EoH, with soft SEDs boosting the temperature fluctuations as heating is confined to the relative proximity of galaxies \citep{Pacucci14}.

The X-ray SED emerging from the host galaxies is determined by (i) the intrinsic X-ray emission of the sources mentioned above (i.e. HMXB, hot ISM, mini-QSO); and (ii) the absorption within the host galaxy. It is very challenging to disentangle these two effects using observations of local X-ray sources at low energies (e.g. $\lsim 1$ keV). Low-energy photons (most relevant for the EoH) can easily be absorbed inside both the host galaxy and the Milky Way, making it difficult to test theoretical models of the intrinsic emission.

{\it In this work we combine the intrinsic SEDs from public stellar population synthesis models of HMXBs, together with ISM absorption estimates from hydrodynamic simulations of the first galaxies, and calculate the resulting 21-cm signal.} The main novelty in our work is to use high-resolution ($\sim$1 comoving pc), cosmological simulations to estimate the absorption within the host galaxies.

Previous EoH studies assume either a relatively ad-hoc ISM absorption (e.g. \citealt{MFS13, MF16}), or use ISM absorption typical of the Milky Way (e.g. \citealt{FBV14}). We expect the first galaxies to be very different from the Milky Way: smaller, more compact, and more susceptible to feedback (e.g. \citealt{FFM15, Pallottini17}).

This paper is organized as follows. In \S \ref{sec:method} we discuss our models for the intrinsic HMXB SEDs (\S \ref{sec:intrisicspec}) and the absorption inside the host galaxy ISM (\S \ref{sec:sim}). In \S \ref{sec:results} we present our results for the emerging X-ray spectrum (\S \ref{sec:emerging}) and the resulting 21-cm forecasts (\S \ref{sec:21cm}). Finally we conclude in \S \ref{sec:conc}. Unless stated otherwise, we quote all quantities in comoving units. Our simulations assume a standard $\Lambda$CDM cosmology, consistent with recent results from the Planck mission \citep{Planck15}.

\section{Modeling the emerging spectrum}\label{sec:method}

To compute the X-ray spectrum emerging from the first galaxies, we populate a high-resolution cosmological simulation with HMXBs (in post-processing), drawing random sight-lines through the simulation to compute the corresponding opacities. Below we discuss the procedure in detail.

The emerging spectrum along a line of sight (LOS) to a star forming system can be written as:
\begin{equation}
 \label{eq:luminosity}
 L(E) = L_0(E) e^{-\tau_{\rm X}(E)} ~ ,
\end{equation}
\noindent where $E$ is the photon energy, $L_0(E)$ the intrinsic spectrum of HMXBs in the system, and $\tau_{\rm X}(E)$ the optical depth along the LOS. We present all spectra in common units of the specific luminosity per unit star formation rate\footnote{Note that since HMXBs correspond to young stellar populations, their combined luminosity output is expected to scale linearly with the galaxy's SFR. This has been confirmed observationally, both in local star-forming galaxies (e.g. \citealt{MGS12_HMXB}) and in moderate redshift galaxies (e.g. \citealt{Lehmer16})}, i.e. erg s$^{-1}$ keV$^{-1}$ $\Msun^{-1}$ yr. 

\subsection{The intrinsic HMXB spectrum from population synthesis}\label{sec:intrisicspec}

\begin{figure}
\vspace{-1\baselineskip}
{
 \includegraphics[width=0.48\textwidth]{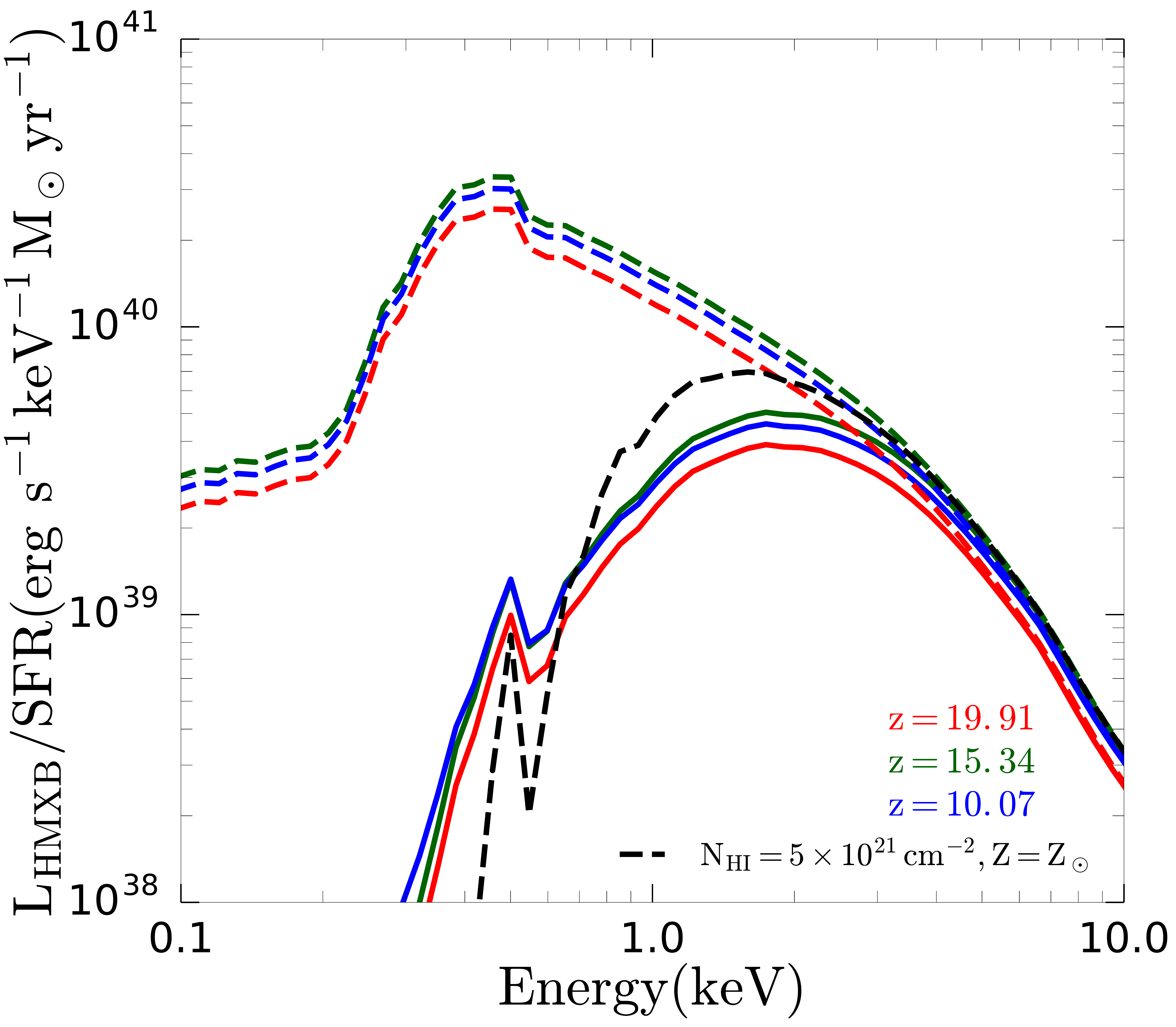}
}
\caption{
The dashed curves correspond to the intrinsic (unabsorbed) HMXB SED, while the solid curves include attenuation typical of local, Milky Way sources (taken from \citealt{Fragos13}, at $z \approx$ 10, 15, 20). The attenuation represented by the solid curves (which was adopted in some prior cosmological 21-cm estimates) is roughly consistent with an HI column density of $N_{\rm HI}\sim 5\times10^{21}$ cm$^{-2}$ assuming solar metallicity ({\it dashed, black curve}).
\label{fig:intrinsicSED}
}
\vspace{-0.5\baselineskip}
\end{figure}

The intrinsic HMXB emission spectrum, $L_0(E)$, is taken from the publicly-available models presented in \citet{Fragos13}\footnote{\url{http://vizier.cfa.harvard.edu/viz-bin/VizieR?-source=J/ApJ/776/L31}}. \citet{Fragos12} performed suites of population synthesis models using StarTrack \citep{Belczynski02, Belczynski08}, calibrated to observations of Galactic neutron star and black hole X-ray binaries at different spectral states. They estimated the intrinsic SEDs by subtracting the assumed ISM absorption from the models used to match observations, and by assuming the intrinsic power-law components do not extend to energies below $\sim$ 1 (0.2) keV in the high (low) state.\footnote{Note that the intrinsic emission and the interstellar and local star system absorption are very degenerate at low energies; thus the \citet{Fragos12} models are very uncertain at energies below 1 keV (T. Fragos, private communication), which unfortunately are the ones relevant for heating the IGM.}. They then used star formation histories from the public Millennium simulation outputs \citep{Boylan09} to predict the X-ray spectral emission from the HMXB populations throughout cosmic time.
%

We show the intrinsic SEDs from \citet{Fragos13} in Fig. \ref{fig:intrinsicSED}. As can be seen from the figure, these SEDs do not have a strong redshift dependence. This is driven by the metal pollution histories in the \citet{Fragos12} models, which predict relatively pristine galaxies ($Z \lsim 0.2 Z_\odot$) beyond $z\gsim8$ (see also \citealt{Pallottini14}). At these low metallicities, the metal-driven stellar winds are inefficient, and population synthesis models converge to the average SEDs shown in Fig. \ref{fig:intrinsicSED} (assuming a fixed stellar initial mass function; \citealt{Belczynski08, MF16}).
For concreteness as well as to simplify the cosmological calculation, we adopt the $z\approx12$ model as our fiducial intrinsic SED, corresponding to the epoch during which X-ray heating is most important for the 21-cm signal (see below).

We normalize the intrinsic SED in the 2-10 keV band, as it is not sensitive to uncertainties in the local absorption (see Fig. \ref{fig:intrinsicSED}). We use the high-redshift limit from \citet[][see their Fig. 3]{MF16}, requiring that the 2-10 keV bolometric luminosity of our SED is $L_{2-10} = 10^{40.5} {\rm erg} ~ {\rm s}^{-1} ~ (\Msun {\rm yr}^{-1})^{-1}$.
This scaling is consistent with the {\it Chandra} 6 Ms observations of moderate-redshift galaxies \citep{Lehmer16}.

\subsection{X-ray opacity from cosmological simulations of the first galaxies}
\label{sec:sim}

\begin{figure*}
\vspace{-1\baselineskip}
{
 \includegraphics[width=0.45\textwidth]{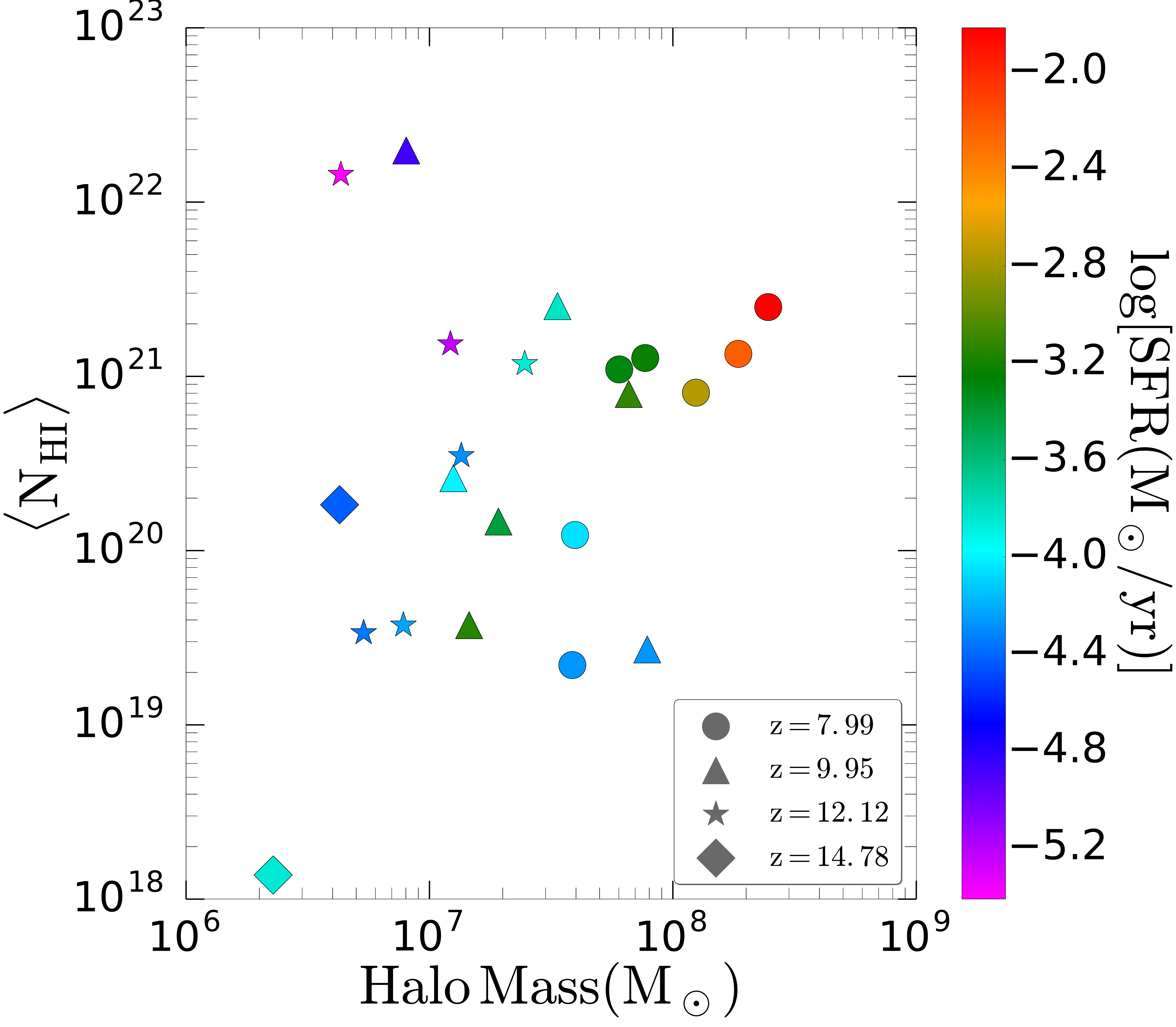}
 \includegraphics[width=0.45\textwidth]{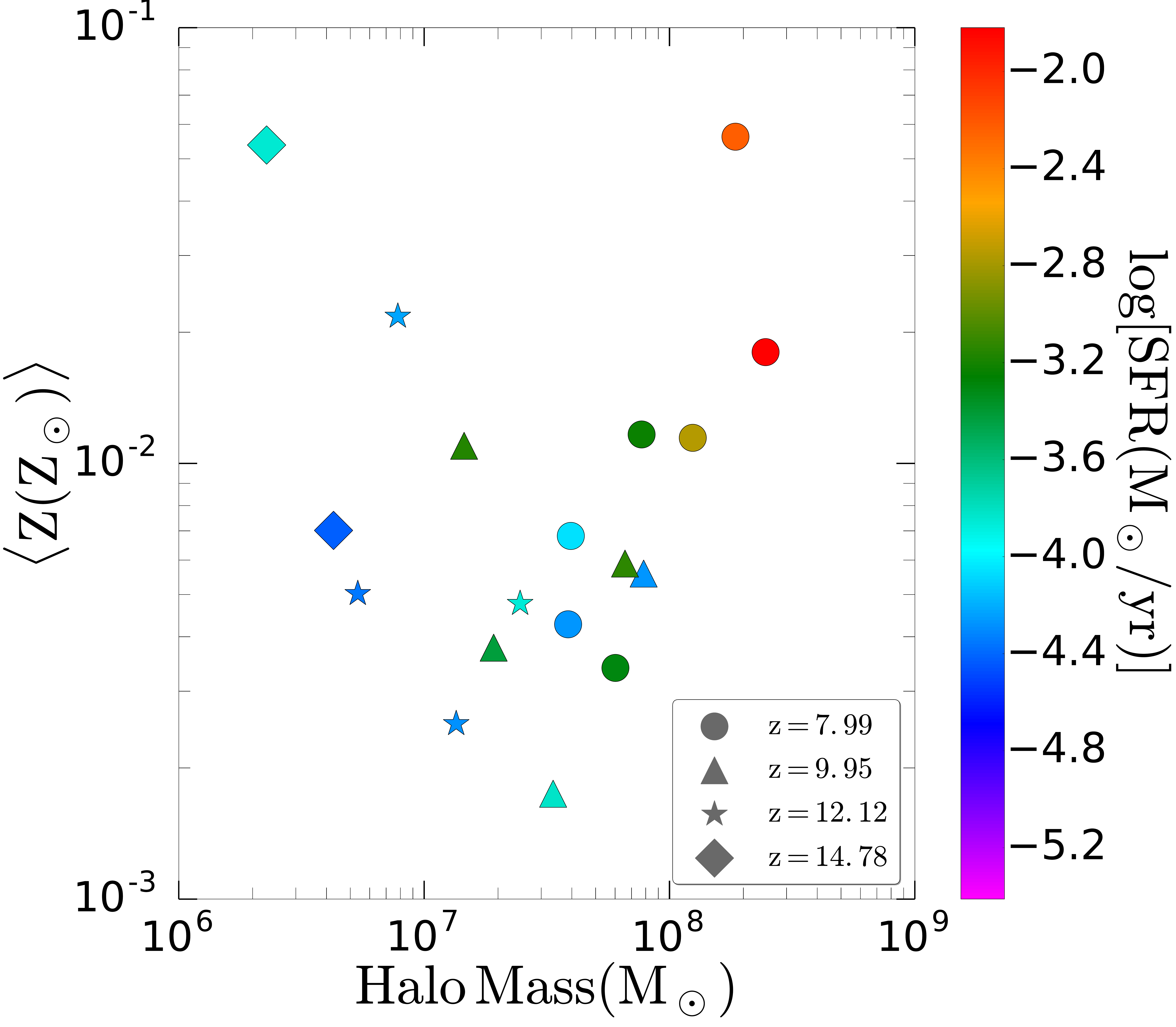}
}
\caption{
{\it Left:} Average HI column densities originating from young star-particles in the simulations of \citet{Wise14}. The averaging is performed over 30000 lines of sight (LOSs) for each star-forming galaxy, each of which is represented by a symbol in the figure. The intrinsic luminosity of HMXBs is taken to scale with the galaxy's SFR (shown with the color bar). {\it Right:} the corresponding density-weighted average metallicity along the LOSs. See text for more details.
\label{fig:column_den}
}
\vspace{-0.5\baselineskip}
\end{figure*}

The optical depth, $\tau_{\rm X}(E)$, along a given LOS depends mostly on the HI column density, $N_{\rm HI}$, and metallicity, $Z$:
\begin{equation}
 \label{eq:tau}
 \tau_{\rm X}(E)= n_{\rm H} \int_0^{R_{\rm vir}} {\rm dr} \left[ ~x_{\rm HI} \sigma_{\rm HI} + \frac{n_{\rm He}}{n_{\rm H}} \sigma_{\rm HeI} + \frac{Z}{Z_\odot} \sigma_{\rm metals} ~\right].
\end{equation}
\noindent Here $n_{\rm H}$ is the hydrogen number density, $x_{\rm HI}$ is the neutral fraction, $n_{\rm He}$ is the helium number density, $R_{\rm vir}$ the virial radius of the host halo, and $\sigma_{\rm HI}(E)$, $\sigma_{\rm HeI}(E)$, $\sigma_{\rm metals}(E)$ are the photo-ionization cross-sections of HI, HeI and metals. The metal cross-section is defined per hydrogen atom and is computed from \citet{MM83} using solar abundance ratios.\footnote{We note that the X-ray opacity is not sensitive to whether or not the metals clump in dust grains (e.g. Fig. 1 in \citealt{MM83}).} For simplicity we conservatively assume all elements heavier than hydrogen are neutral.\footnote{Although helium photo-ionization is not tracked in the simulation, we roughly test the impact of helium ionization of the mean X-ray opacity by adopting various temperature thresholds above which we assume helium is single and doubly ionized: few$\times10^4$ K for HeII and few$\times10^5$K for HeIII. The resulting opacities are almost indistinguishable with our fiducial model, as the bulk of the absorption stems from denser, neutral gas.}

We take the hydrogen number density, $n_{\rm H}$, neutral fraction, $x_{\rm HI}$, and metallicity, $Z$, from the cosmological radiation hydrodynamics simulation of \citet{Wise14}, computed using the adaptive mesh refinement code {\sc Enzo} \citep{Bryan13} and its radiation transport module {\sc Moray} \citep{Wise11}. This 1 Mpc on a side, ``first galaxy'' simulation follows the buildup of 32 galaxies, including the effects of radiative and supernova feedback from both massive, metal-free (Population III) and metal-enriched stars. For this work, we focus on the $\sim$ 7 galaxies which are actively forming stars in any given snapshot, and are thus expected to host active HMXBs. Other relevant physical processes included in the simulation are radiative cooling from metals, direct radiation pressure from starlight, and a time-dependent Lyman-Werner (H$_2$ dissociating) radiation background. The simulation has a maximal spatial resolution of 1~comoving parsec, dark matter particle mass of 1840~$M_\odot$, and is halted at $z = 7.3$. At this time, the most massive galaxy has a stellar mass of $3.7 \times 10^6 M_\odot$ hosted in a halo of mass $6.8 \times 10^8 M_\odot$.

For each galaxy, we randomly shoot sight-lines originating from those star particles which are younger than 20 Myr (and are thus expected to host HMXBs; e.g. \citealt{Fragos12}). We integrate eq. \ref{eq:tau} out to the halo virial radius of each halo, $R_{\rm vir}$. We sample each galaxy with 30000 LOSs, resulting in at least 1000 LOSs per active star particle.

In the left panel of Fig. \ref{fig:column_den} we show the average HI column densities for star forming halos. The color bar represents the total star formation rate of the galaxy. The figure illustrates halos becoming more massive with higher star formation rates as time progresses. Star formation is dominated by halos with masses around $\gsim10^8 \Msun$, whose virial temperature is high enough for baryons to be able to efficiently cool through atomic hydrogen transitions. For these halos the average column density\footnote{We confirm that the ionizing escape fraction of these halos is a few percent, as found by \citet{Wise14}. This average ionizing photon escape fraction is however driven by a few rare low-column density sight-lines, while the bulk of the sight-lines have an escape fraction of $\sim$zero. Thus the average column density shown in the left panel of figure \ref{fig:column_den} is not a good proxy for the average escape fraction. To put it differently: $\langle \exp[-\tau_{\rm uv}] \rangle \neq \exp[\langle -\tau_{\rm uv} \rangle]$. In contrast, the X-ray absorption (at most of the relevant energies) is {\it not} zero. Therefore, the rare tail of low-column density sight-lines is less important for the average X-ray opacity, and one could reasonably use the average column density when estimating the average X-ray opacity, i.e. $\langle \exp[-\tau_{\rm X}] \rangle \sim \exp[\langle -\tau_{\rm X} \rangle]$. This is explicitly seen in Fig. \ref{fig:opacity}.} is around 10$^{21}$--10$^{21.5}$ cm$^{-2}$.

We note that our simulated column densities are somewhat lower than absorption seen in high-$z$ gamma ray bursts (GRBs; e.g. \citealt{Totani06, Campana12, Salvaterra15}) afterglow spectra. However, estimating the column densities from optical or X-ray GRB spectra is very challenging, as there is a degeneracy between the intrinsic GRB + ISM emission, the host ISM absorption, as well as the intervening absorption from the IGM and/or the Milky Way. Indeed, \citet{Campana15} find that the high column densities infered from X-ray GRB spectra at high-$z$ can be entirely explained with intervening systems.
%

In the right panel of Fig. \ref{fig:column_den}, we show the average metallicity of the galaxies, weighted by the density along the LOS. In addition to the HI column densities shown in left panel, the metallicity sets the X-ray absorption of the host halo. Here we do see a marked difference from Milky Way-like absorption, typically assumed in the literature. Since these galaxies are relatively pristine, we expect the contribution of metals to the X-ray optical depth to be sub-dominant to that of hydrogen and helium. Indeed, all of the star-forming galaxies in our sample have density-weighted metallicities below 0.06 $Z_\odot$.

\section{Results}
\label{sec:results}

Following the procedure outlined above, we now present the resulting X-ray spectra emerging from our $z>8$ galaxies in \S \ref{sec:emerging}. Then in \S \ref{sec:21cm} we present the corresponding forecasts for the cosmic 21cm signal, adopting various SFR histories.

\subsection{The emerging spectrum}
\label{sec:emerging}

\begin{figure}
\vspace{-1\baselineskip}
{
 \includegraphics[width=0.5\textwidth]{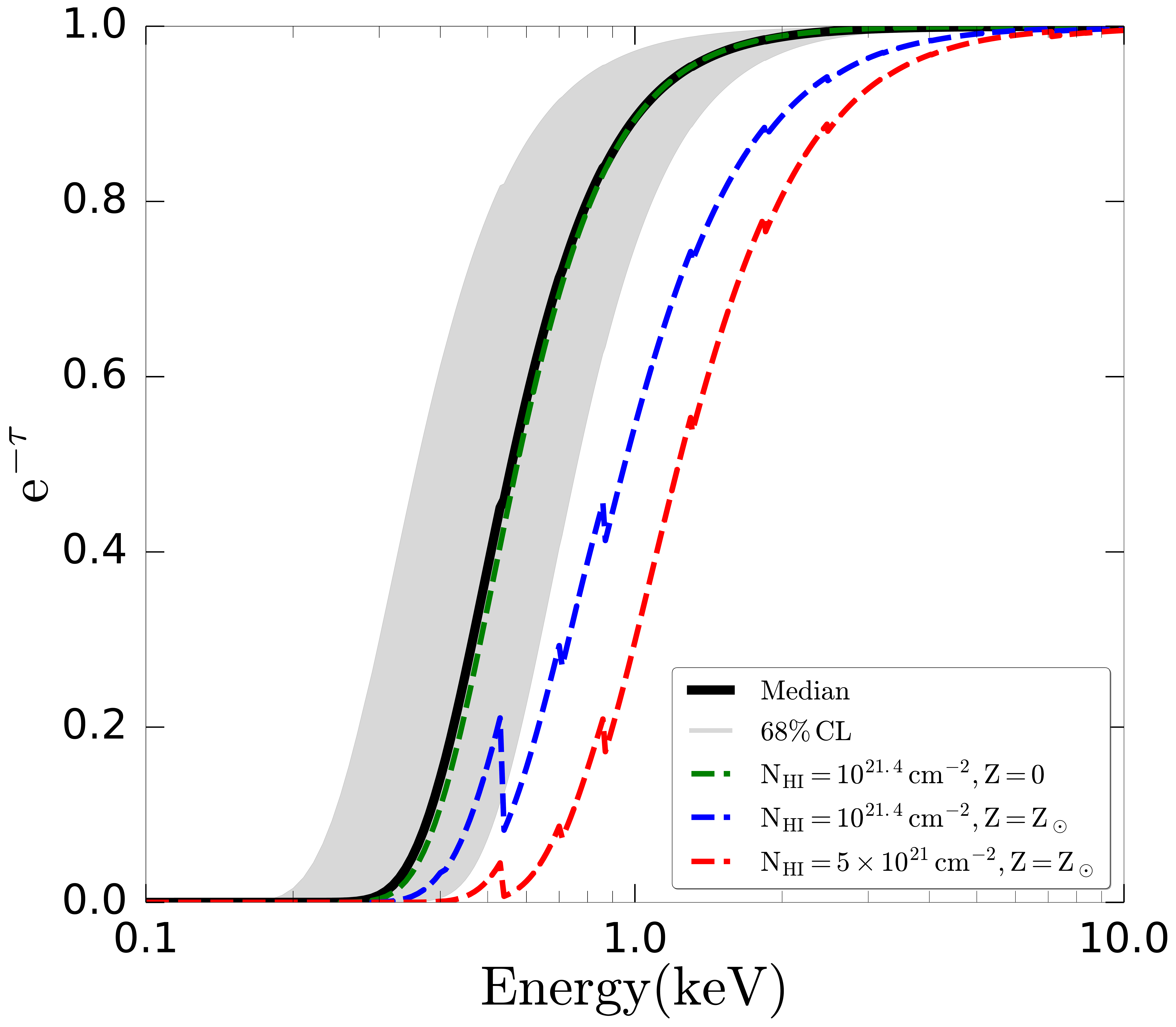}
}
\caption{
Opacity through the $z>8$ galaxies in our simulations. We weigh the opacity from each LOS with the host galaxy's SFR (taken to be proportional to the HMXB luminosity). We sample all of the 22 star-forming galaxy snapshots shown in Fig. \ref{fig:column_den}, though our weighing biases our results to the atomically-cooled galaxies (in halos of mass $\gsim10^8 \Msun$), which have the most efficient star formation.
The median of the resulting SFR-weighted distribution is shown with a black solid curve, while the 68\% C.L. are shown with the shaded region. We find that an opacity corresponding to a $N_{\rm HI}=10^{21.4}$ cm$^{-2}$, metal-free ISM ({\it green dashed curve}) matches the median from our simulations very well.
Likewise, we find that the 68\% C.L. are well reproduced with a metal-free ISM with $N_{\rm HI}=10^{21.80}$ cm$^{-2}$ and $N_{\rm HI}=10^{20.75}$ cm$^{-2}$ (not explicitly shown so as not to overcrowd the figure).
The blue dashed curve highlights the impact of the metallicity, while the red dashed curve corresponds roughly to the opacity to local, Milky Way sources (cf. Fig. \ref{fig:intrinsicSED}).
\label{fig:opacity}
}
\vspace{-0.5\baselineskip}
\end{figure}

In Fig. \ref{fig:opacity} we show the median opacity through our star-forming galaxies ({\it black curve}), and the 68\% confidence limits (C.L.) ({\it shaded region}). Since the luminosity of the HMXB population scales linearly with the SFR (e.g. \citealt{Fragos12, MGS12_HMXB}), we weigh the opacity from each LOS with the SFR of the host galaxy.\footnote{Note that in doing so, our opacity distributions are biased towards the actively star-forming galaxies, which in general will have higher column densities than the average galaxy. However our goal in this work is to predict the X-ray background from HMXB, whose luminosity scales with the SFR of the host galaxy; therefore weighing the opacity by the SFR is appropriate for our purposes.} We do not differentiate between redshift outputs, sampling all of the star-forming galaxies shown in the previous two figures. This allows us to increase the statistical sample, without any obvious biases (there are no clear redshift trends in Fig. \ref{fig:column_den} when comparing at a fixed halo mass). We also note that \citet{Xu16} show that the simulations used here are indeed statistically representative of a larger sample.

Fig. \ref{fig:opacity} illustrates one of the main findings of this work: {\it the X-ray opacity of the galaxies driving the epoch of cosmic heating can be approximated by a metal-free ISM with a typical column density of $\log[N_{\rm HI}/{\rm cm^{-2}}] = 21.4^{+0.40}_{-0.65}$.}. The quoted limits enclose 68\% of the uncertainty in the opacity.
Due to the relative dearth of metals, the ISM in high-$z$ galaxies is more transparent than in local galaxies. Thus photons with energies as low as 0.5 keV can escape into the IGM. Moreover, the resulting $\exp[-\tau_{\rm X}]$ is a much steeper function of energy than if metals contribute significantly.
Our findings therefore support relatively ad-hoc assumptions used in prior works (e.g. \citealt{Furlanetto06, PF07, MFS13}) for a sharp cut-off in energy, roughly corresponding to these column densities in a metal poor environment.

\begin{figure}
\vspace{-1\baselineskip}
{
 \includegraphics[width=0.5\textwidth]{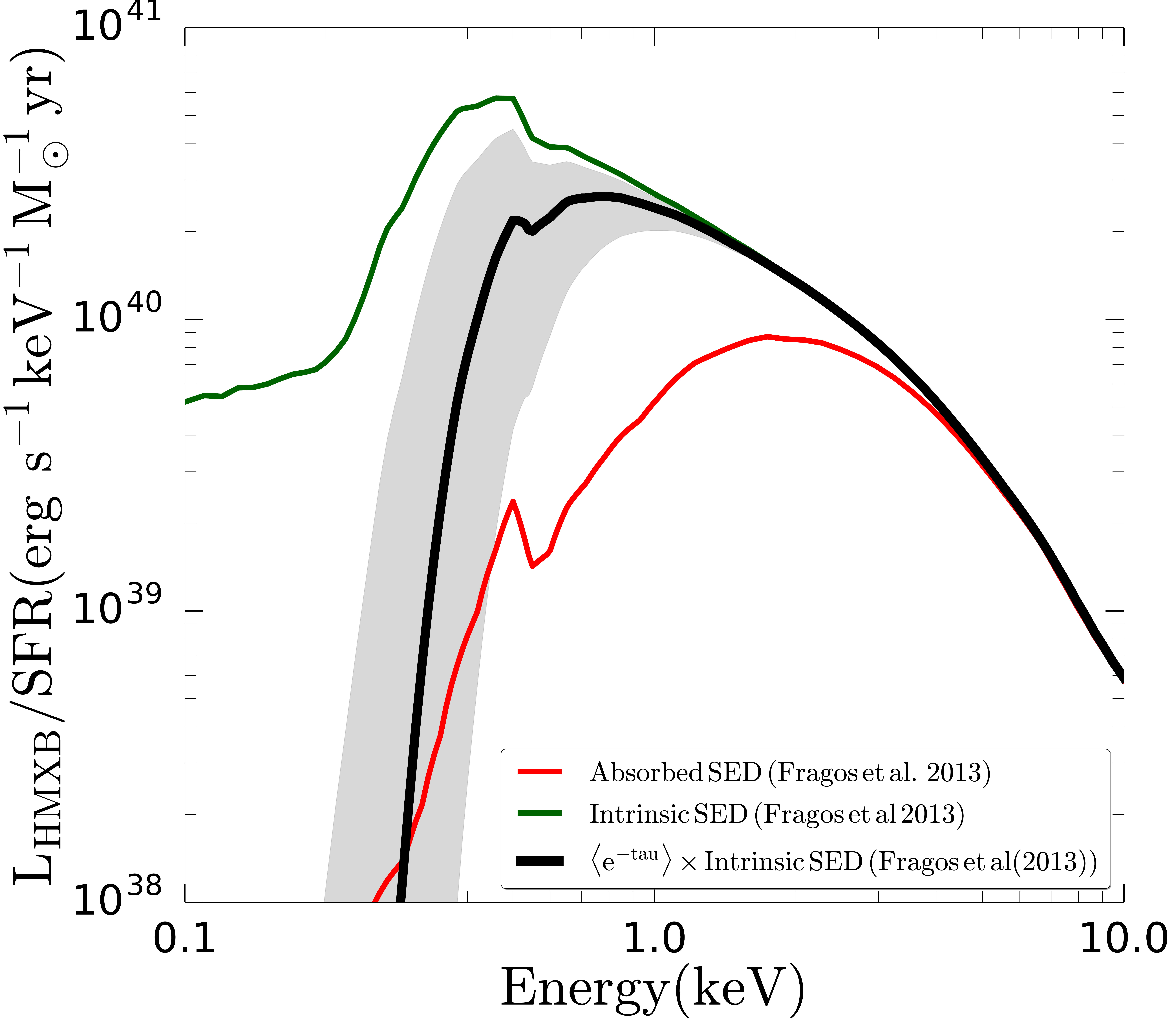}
}
\caption{
Intrinsic (absorbed) HMXB SEDs from \citet{Fragos13} are shown with the green (red) curves. The back curve instead uses the median SFR-weighted opacities from our simulations (shown in Fig. \ref{fig:opacity}). As in the Fig. \ref{fig:opacity}, the shaded region shows the 68\% C.L. on the opacity.
}
\label{fig:SED}
\vspace{-0.5\baselineskip}
\end{figure}

In Fig. \ref{fig:SED} we combine these opacities with the intrinsic HMXB emission, showing the X-ray spectrum emerging from the galaxies into the IGM ({\it black curve}). The interstellar absorption from our simulations results in only a few percent of $\sim$ 1 keV photons being absorbed within the host galaxies. This is in stark contrast with the red curve showing the previously-used absorption model (motivated by local observations). This absorption results in a factor of $\sim5$ fewer 1 keV photons escaping the galaxy.

Overall, the soft-band ($<2$ keV) luminosity of HMXBs which is responsible for IGM heating is $1.8\times10^{40}$ erg s$^{-1}$ $\Msun$ yr$^{-1}$ for our median opacity. This is over a factor of three larger than the soft-band luminosity corresponding to the red curve, $5.3\times10^{39}$ erg s$^{-1}$ $\Msun$ yr$^{-1}$. This comparably low soft-band luminosity largely explains the inefficient X-ray heating suggested by \citet[][who used the SED shown with the red curve]{FBV14}, as we quantify further below.

\subsection{The cosmic 21-cm signal}
\label{sec:21cm}

We now make forecasts for the cosmic 21-cm signal, corresponding to the X-ray SEDs shown in the previous section. The 21-cm signal is usually represented in terms of the offset of the 21-cm brightness temperature from the CMB temperature, $\Tcmb$.  Our simulated 21-cm signal is a function of both redshift, $z$, and spatial position, ${\bf x}$:
\begin{align}
\label{eq:delT}
\nonumber \delT({\bf x}, z) \approx &27 \nf (1+\delNL) \left(\frac{H}{dv_r/dr + H}\right) \left(1 - \frac{\Tcmb}{\Ts} \right) \\
&\times \left( \frac{1+z}{10} \frac{0.15}{\Omega_{\rm M} h^2}\right)^{1/2} \left( \frac{\Omega_b h^2}{0.023} \right) {\rm mK} .
\end{align}
\noindent Here $\Ts$ is the gas spin temperature, $\delNL({\bf x}, z) \equiv \rho/\bar{\rho} - 1$ is the evolved (Eulerian) density contrast, $H(z)$ is the Hubble parameter, $dv_r/dr$ is the comoving gradient of the line of sight component of the comoving velocity. From the above, we see that the EoR can be studied through the $\nf$ factor, while the EoH is identified through the spin temperature. The spin temperature interpolates between the CMB temperature and the gas kinetic temperature, $T_K$. Since the observation uses the CMB as a back-light, a signal is only obtained if $T_S$ approaches $T_K$. For the epoch of interest here, this coupling is achieved through a Lyman alpha background [so-called Wouthuysen-Field (WF) coupling; \citealt{Wouthuysen52, Field58}].

To compute $\delT({\bf x}, z)$, we use the public \cmfast\ code\footnote{\url{https://github.com/andreimesinger/21cmFAST}} \citep{MF07, MFC11}. Our simulations are 500 Mpc on a side, with a 400$^3$ grid. This cosmological ``semi-numerical'' simulation uses perturbation theory to generate the density and velocity fields \citep{Zeldovich70}. UV ionizations are computed with a modified version of the photon counting approach of \citet{FHZ04}: comparing the (cumulative) number of ionizing photons to the number of neutral atoms plus recombinations inside spheres of decreasing radii. The X-ray and Lyman alpha backgrounds, driven by photons with much longer mean free paths, are computed by integrating along the light-cone.

As the focus in this work is the heating resulting from our X-ray SEDs, we do not vary the astrophysical parameters which determine WF coupling and the EoR. As seen below, our fiducial choices result in EoR histories which are consistent with the latest observations (for a detailed summary of current EoR constraints, see \citealt{GM16}).

The angle-averaged specific X-ray intensity (in erg s$^{-1}$ keV$^{-1}$ cm$^{-2}$ sr$^{-1}$), which drives the EoH is computed by integrating along the light-cone (cf. \citealt{HM96}):
\begin{equation}
 J(E, {\bf x}, z)=\frac{(1+z)^3}{4\pi}\int_z^{\infty}dz^\prime\frac{c dt}{dz^\prime}\epsilon_{\rm X} ~ ,
\label{eq:J}
\end{equation}
\noindent with the comoving specific emissivity evaluated at the rest-frame energy $E_{\rm e}= E (1+z^\prime)/(1+z)$:
\begin{align}
 \label{eq:emissivity}
 \epsilon_{\rm X}(E_{\rm e}, {\bf x}, z^\prime)& = L(E_{\rm e}) ~ \left[ \rho_{\rm crit,0} \Omega_b f_\ast (1+\delta_{\rm nl}) \frac{df_{\rm coll}}{dt} \right] ~ .
\end{align}
\noindent Here, $L(E_{\rm e})$ is the X-ray spectrum emerging from galaxies (eq. \ref{eq:luminosity}), $\Omega_m \rho_{\rm crit,0}$ is the comoving matter density, $f_{\rm coll}$ is the fraction of matter inside halos, and $f_\ast$ is the fraction of the halo baryons in stars. The quantity in the brackets is the star formation rate density (SFRD) along the light-cone. Our models for the SFRD assume a constant $f_\ast=0.05$ for halos above some critical threshold virial temperature required for efficient star formation, $\Tvirmin$. With this assumption, the SFRD shown in eq. \ref{eq:emissivity} can be expressed as a function of the time derivative of the collapse fraction, $f_{\rm coll}(>\Tvirmin, {\bf x}, z^\prime)$, simplifying the calculation. Although crude, our assumption of a step function $f_\ast$ is sufficient to show general trends.

\begin{figure}
\vspace{-1\baselineskip}
{
 \includegraphics[width=0.5\textwidth]{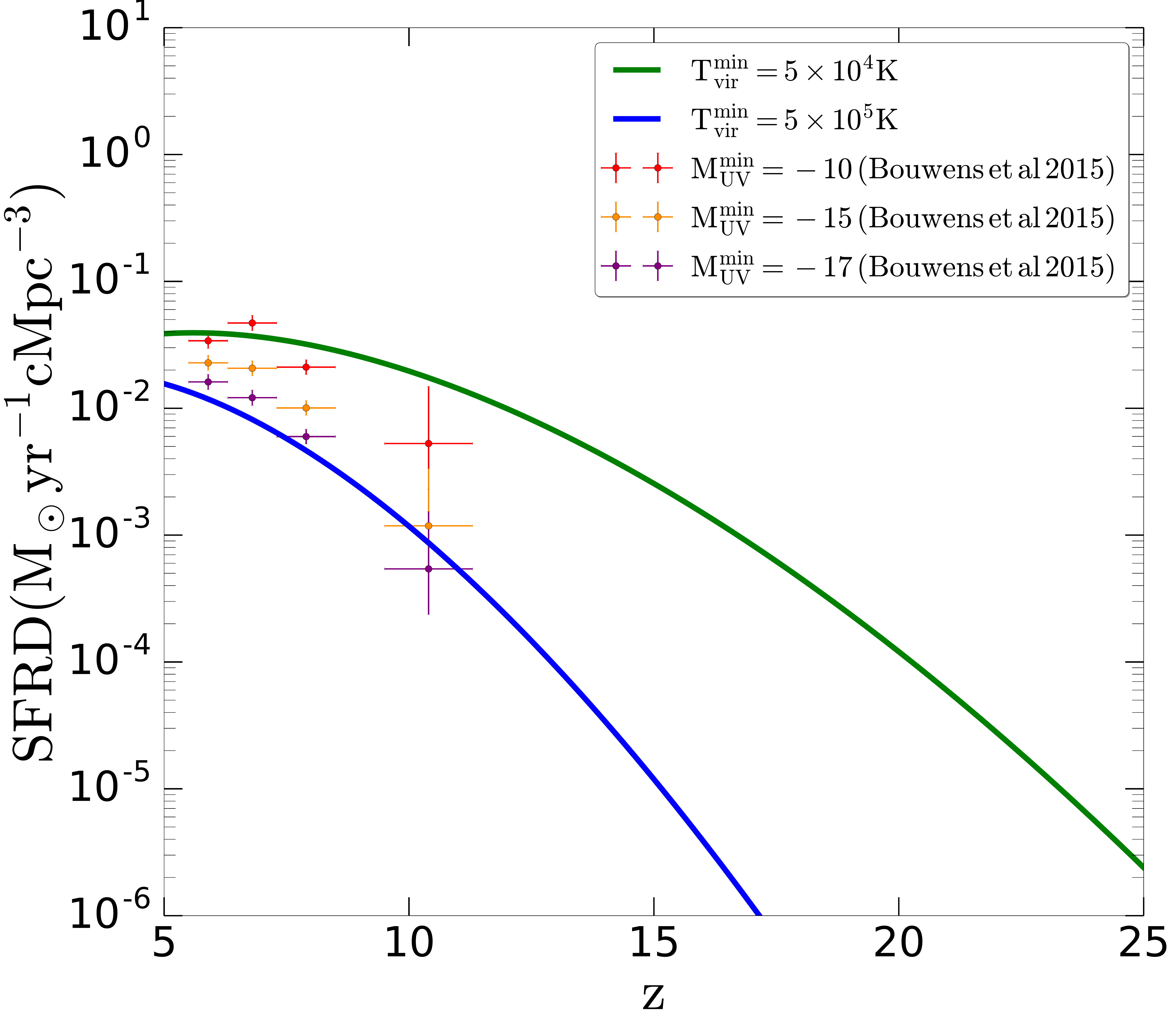}
}
\caption{
Evolution of the global star formation rate density (SFRD). Points correspond to observational estimates from \citet{Bouwens15}, adopting various values for the limiting magnitude, $M_{\rm UV}^{\rm min}$.
}
\label{fig:SFRD}
\vspace{-0.5\baselineskip}
\end{figure}

\begin{figure}
\vspace{-1\baselineskip}
{
 \includegraphics[width=0.5\textwidth]{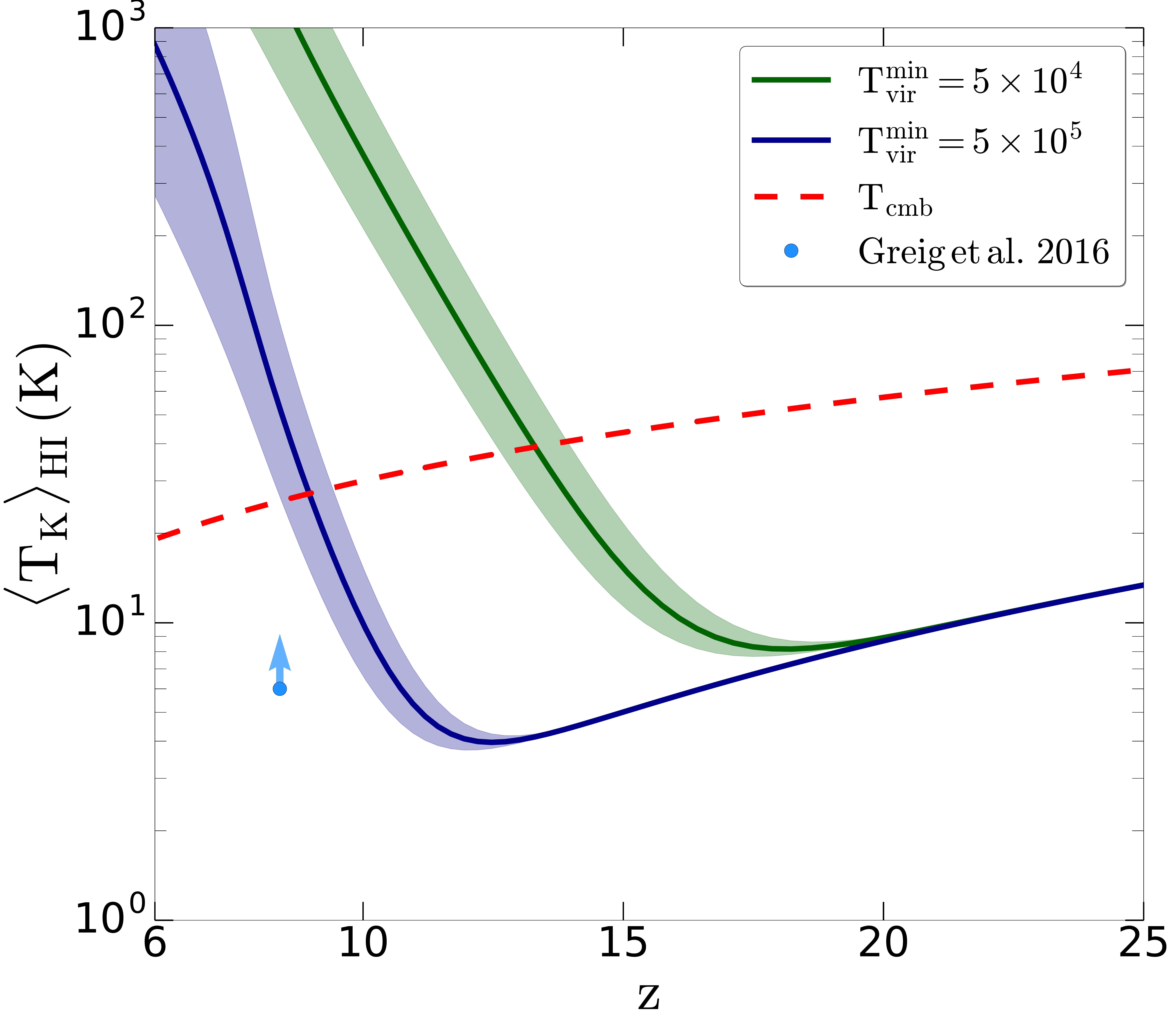}
}
\caption{
 Evolution of the mean temperature inside the predominantly neutral IGM, corresponding to the two SFRD histories in Fig. \ref{fig:SFRD}. The lines and shaded regions correspond to the median and 68\% C.L. of our computed X-ray SEDs (cf. Fig. \ref{fig:SED}). We also include the 1-$\sigma$ lower limit from \citet{GMP16}, derived from recent 21-cm observations.
\label{fig:temp}
}
\vspace{-0.5\baselineskip}
\end{figure}

\begin{figure*}
\vspace{-1\baselineskip}
{
 \includegraphics[width=0.45\textwidth]{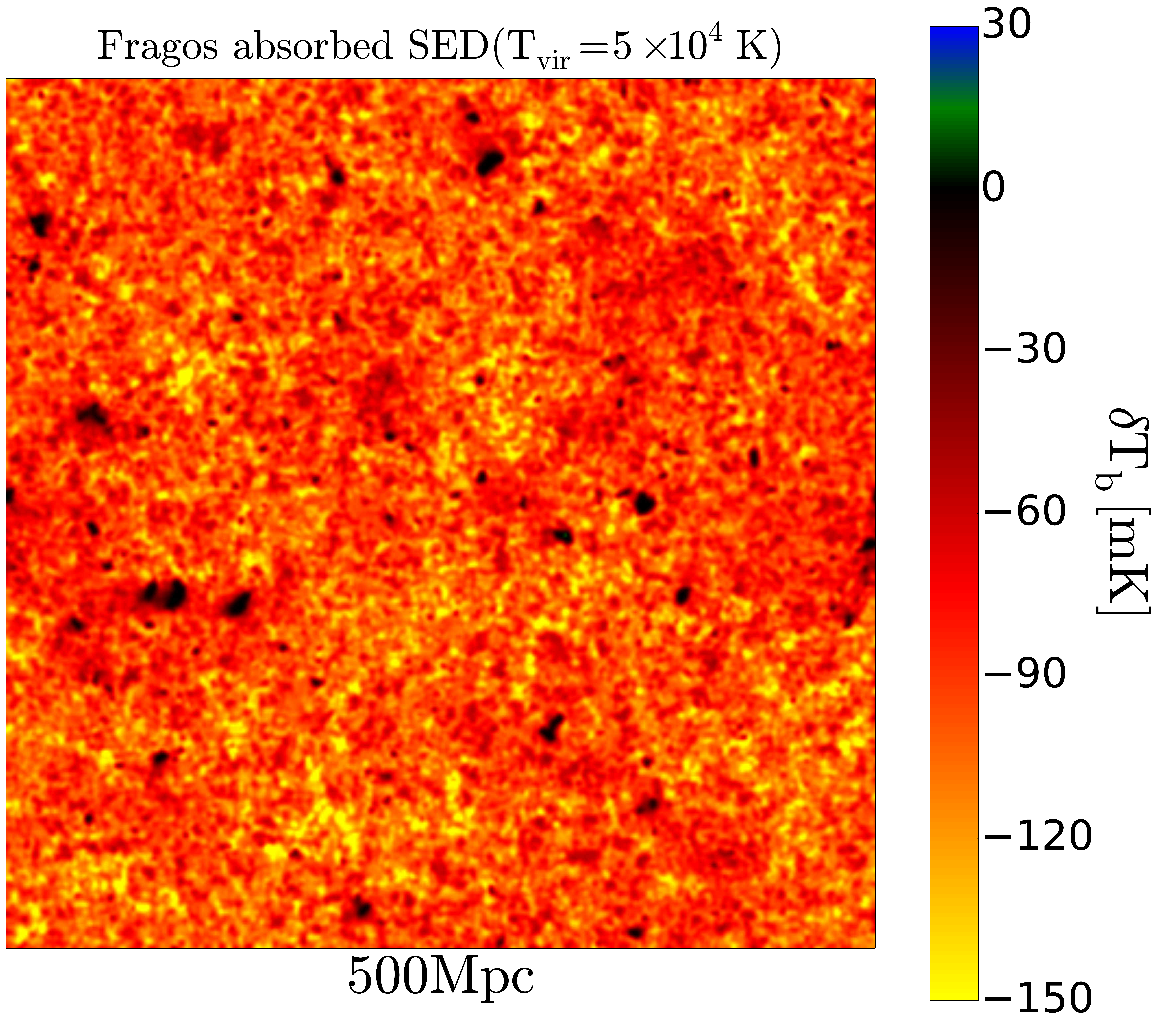}
 \includegraphics[width=0.45\textwidth]{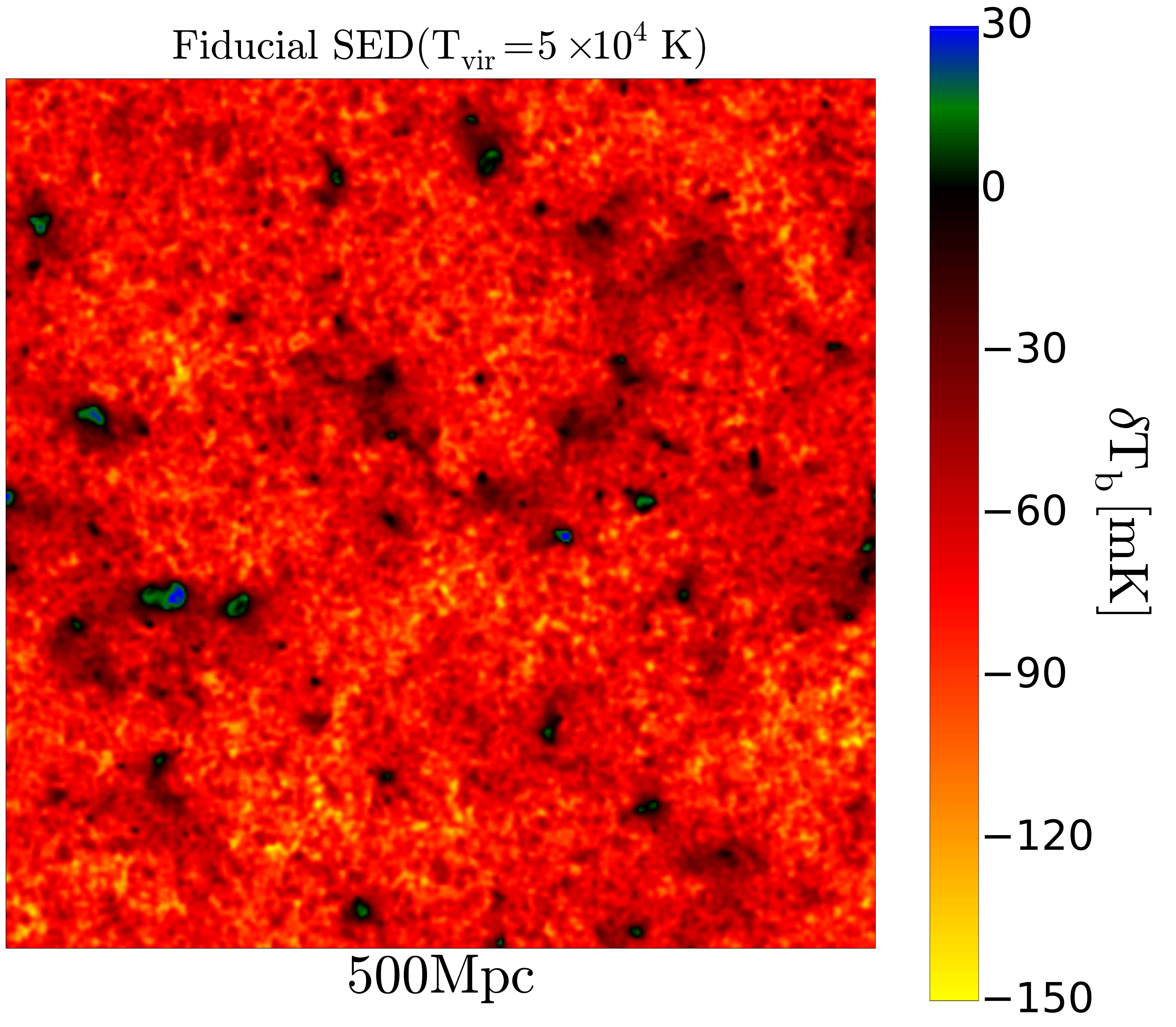}
}
\caption{
Slices through the 21-cm brightness temperature fields. Slices are 500 Mpc on a side and 1.25 Mpc deep. Both assume our fiducial SFRD evolution (i.e. green curve in the previous two figures). The left panel was computed using the absorption from \citet[][motivated by local HMXBs]{Fragos13}, while the right panel corresponds to our fiducial SED computed from our simulations. The slices are taken at $z_{\rm peak}=15.1$ (fiducial SED) and $z_{\rm peak}=13.7$ (absorbed SED from \citealt{Fragos13}), when the large-scale ($k=0.1$ Mpc$^{-1}$) 21-cm power is at its peak, driven by spatial fluctuations in the gas temperature.
\label{fig:slice}
}
\vspace{-0.5\baselineskip}
\end{figure*}

The cosmic SFR history is very uncertain at high redshifts, since we cannot detect the faintest galaxies which could be the dominant population.
Therefore here we compute the SFR history using two values for the limiting halo virial temperature: $\Tvirmin= 5\times10^4$ and $5\times10^5$ K. The corresponding SFRDs\footnote{We note that our approach is not completely self-consistent: the hydrodynamic simulations used to compute the HMXB opacities are not also used to predict the SFRD and the corresponding halo-SFR relations. We keep the global SFR calculation independent so as to allow our forecasts to be more flexible. We note that the 21-cm signal is driven by fluctuations on scales of tens--hundreds of Mpc; the relevant fluctuations of the SFR and halo-SFR relations are unlikely to be well captured by a single 1 Mpc hydro simulation. Indeed as seen from Fig. 2, these simulations do not have any halos with mass $\gsim10^{8.5} \Msun$ at these redshifts.} are shown in Fig. \ref{fig:SFRD}. Also shown are the observational estimates from \citet{Bouwens15}, obtained by integrating down the faint end slope of the observed UV luminosity functions to various limiting magnitudes, $M_{\rm UV}^{\rm min}$. For context, note that $M_{\rm UV}^{\rm min} = -17$ corresponds roughly to the observed limit using unlensed deep fields with the {\it Hubble} telescope. However, gravitational lensing magnification provided by the Frontier Fields clusters has yielded even fainter galaxies, with the faint end slope shown to extend down to at least $M_{\rm UV}^{\rm min} =$ -15 to -13 at $z\sim6$--7 \citep{Atek15, LFL16, Bouwens16}. Although the uncertainties become large at high magnifications, these observations motivate our $\Tvirmin= 5\times10^4$ K model ({\it green curve}) as being more realistic than the $\Tvirmin= 5\times10^5$ K one ({\it blue curve}). Hence, below we shall refer to the $\Tvirmin= 5\times10^4$ K model as ``fiducial''.

In Fig. \ref{fig:temp} we show the evolution of the volume-weighted mean temperature inside the predominately neutral IGM (i.e. outside of the cosmic ionized patches), corresponding to these two SFRD histories. The lines and shaded regions correspond to the median and 68\% C.L. of our computed X-ray SEDs (cf. Fig. \ref{fig:SED}). We also show the 1-$\sigma$ lower limit for this quantity from \citet{GMP16}, derived from a combination of observations: (i) the upper limit on the 21-cm power spectrum at $z=8.4$ measured by PAPER-64 \citep{Ali15}; (ii) the dark fraction in the Lyman forests of high-$z$ QSO spectra \citep{MMO15}; (iii) the Thompson scattering optical depth to the CMB \citep{Planck15}.

It is evident from the figure that the uncertainty in the cosmic SFRD evolution is more important than the comparably minor scatter in the X-ray opacities of our simulated galaxies. The blue curve, corresponding to our extreme model with $\Tvirmin=5\times10^5$ K, results in the predominantly neutral IGM being heated above the CMB relatively late at $z\sim9$, when roughly half of the volume of the Universe is already ionized. This evolution is consistent with the result in \citet{MF16}, whose fiducial model only accounts for star formation in galaxies down to $M_{\rm UV}^{\rm min}\sim -16$.

As discussed above, the green curve is likely more realistic as it accounts for star formation inside fainter galaxies, whose presence at $z\sim6$ has already been detected using cluster lensing \citep{Atek15, LFL16, Bouwens16}.
In this model, the Universe is heated earlier, at $z\sim13$. This result is consistent with most prior studies of the EoH, when accounting for the slightly lower star formation efficiencies and updated $L_X$/SFR relations used here (e.g. \citealt{Furlanetto06, PF06, MGS16}).

We now present the complete 21-cm signal from our simulations (see eq. \ref{eq:delT}).
In Fig. \ref{fig:slice} we show slices through the 21-cm maps, corresponding to the epoch in which temperature fluctuations dominate the signal.
The figure illustrates that our fiducial X-ray SED results in a more inhomogeneous temperature field, compared with the \citet{Fragos13} absorption (which was motivated by local HMXBs). These spatial fluctuations in the gas temperature are driven by the soft X-ray photons, which are more abundant in our fiducial SED. Since the mean free path of the X-rays scales roughly as $\propto E^3$, these soft photons imprint the IGM temperature fluctuations (see \citealt{Pacucci14}). Assuming absorption typical of the Milky Way washes-out the fluctuations and makes heating less efficient (cf. \citealt{FBV14}).

\begin{figure*}
\vspace{-1\baselineskip}
{
 \includegraphics[width=0.45\textwidth]{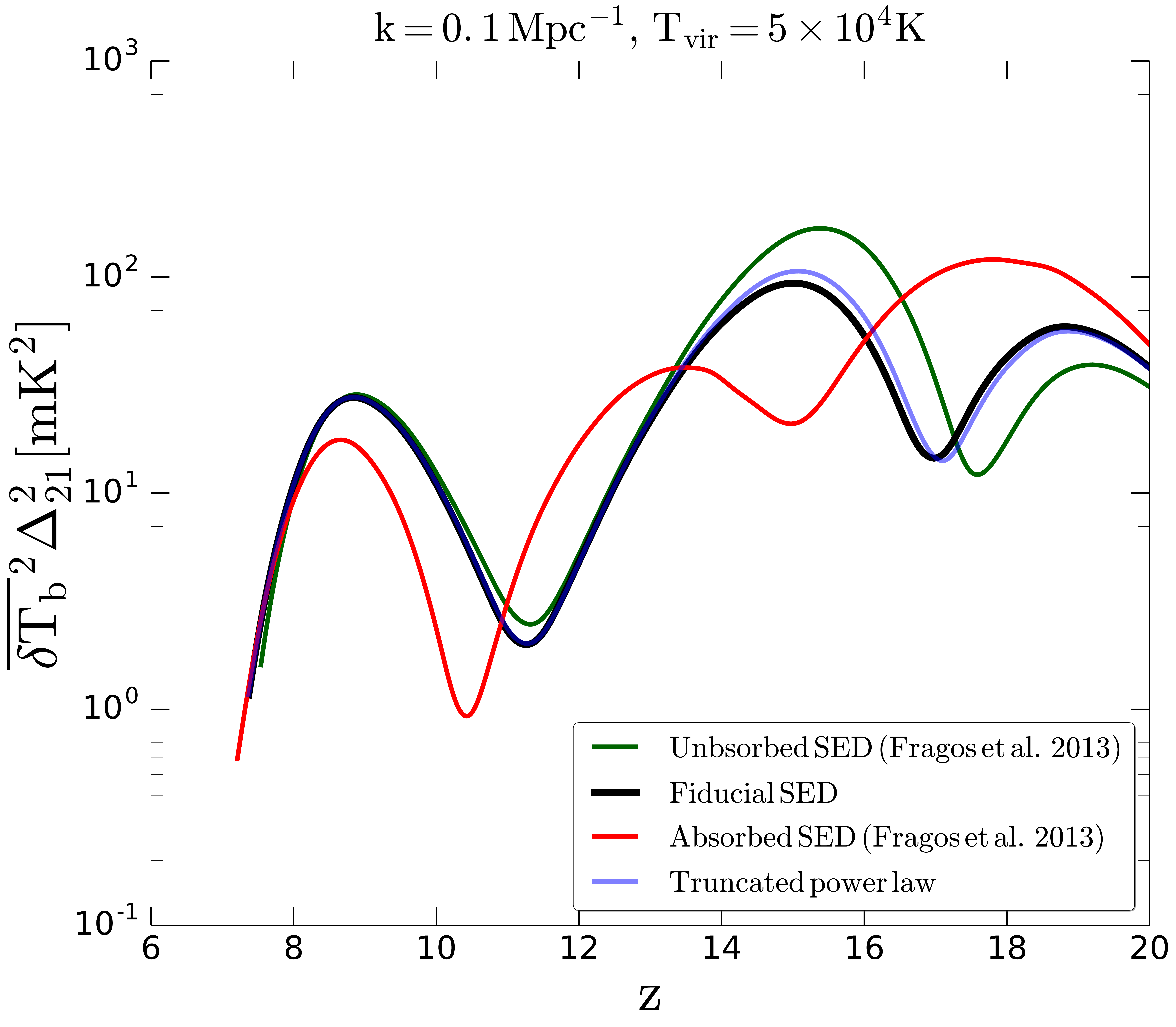}
 \includegraphics[width=0.45\textwidth]{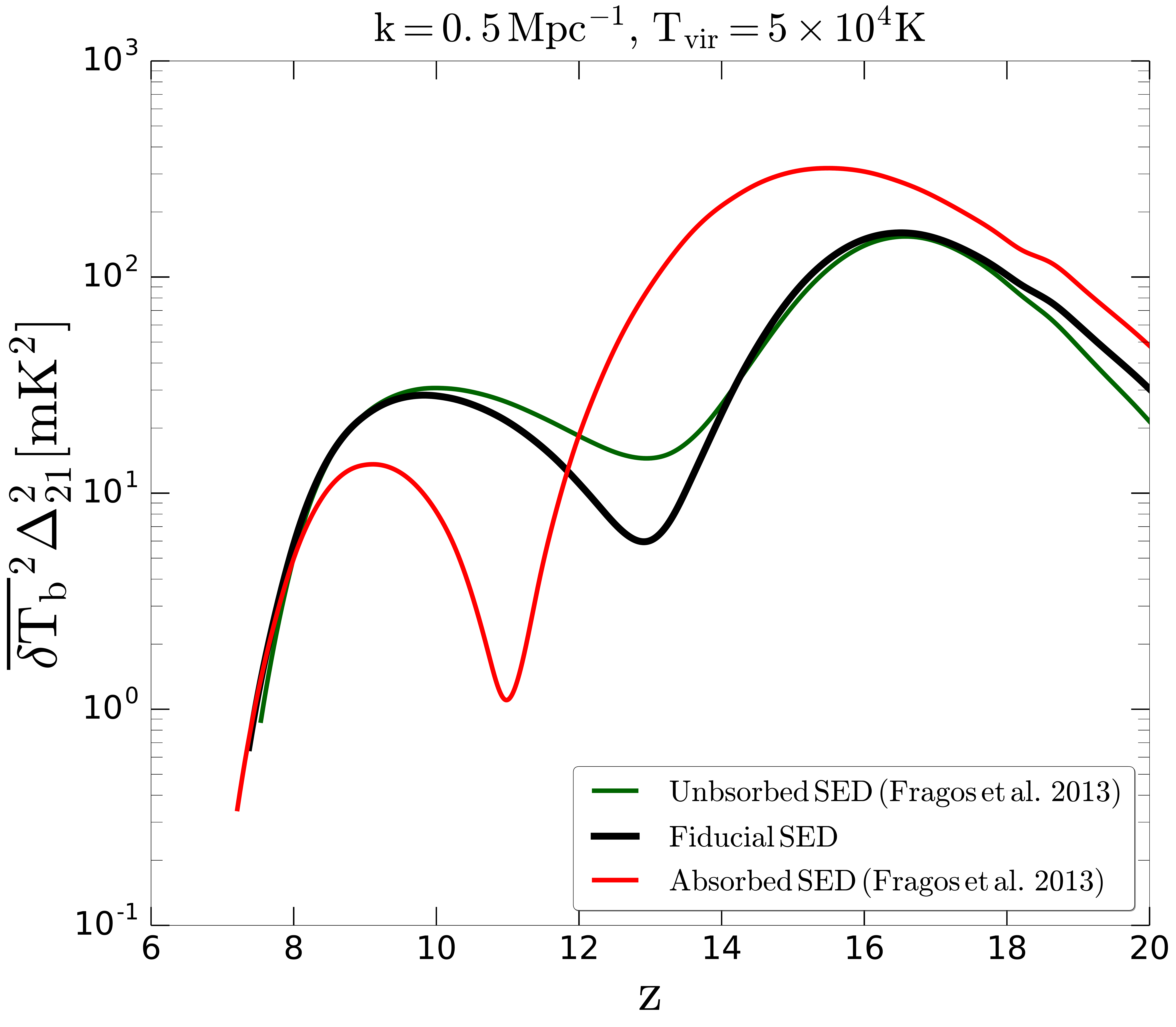}
 \includegraphics[width=0.45\textwidth]{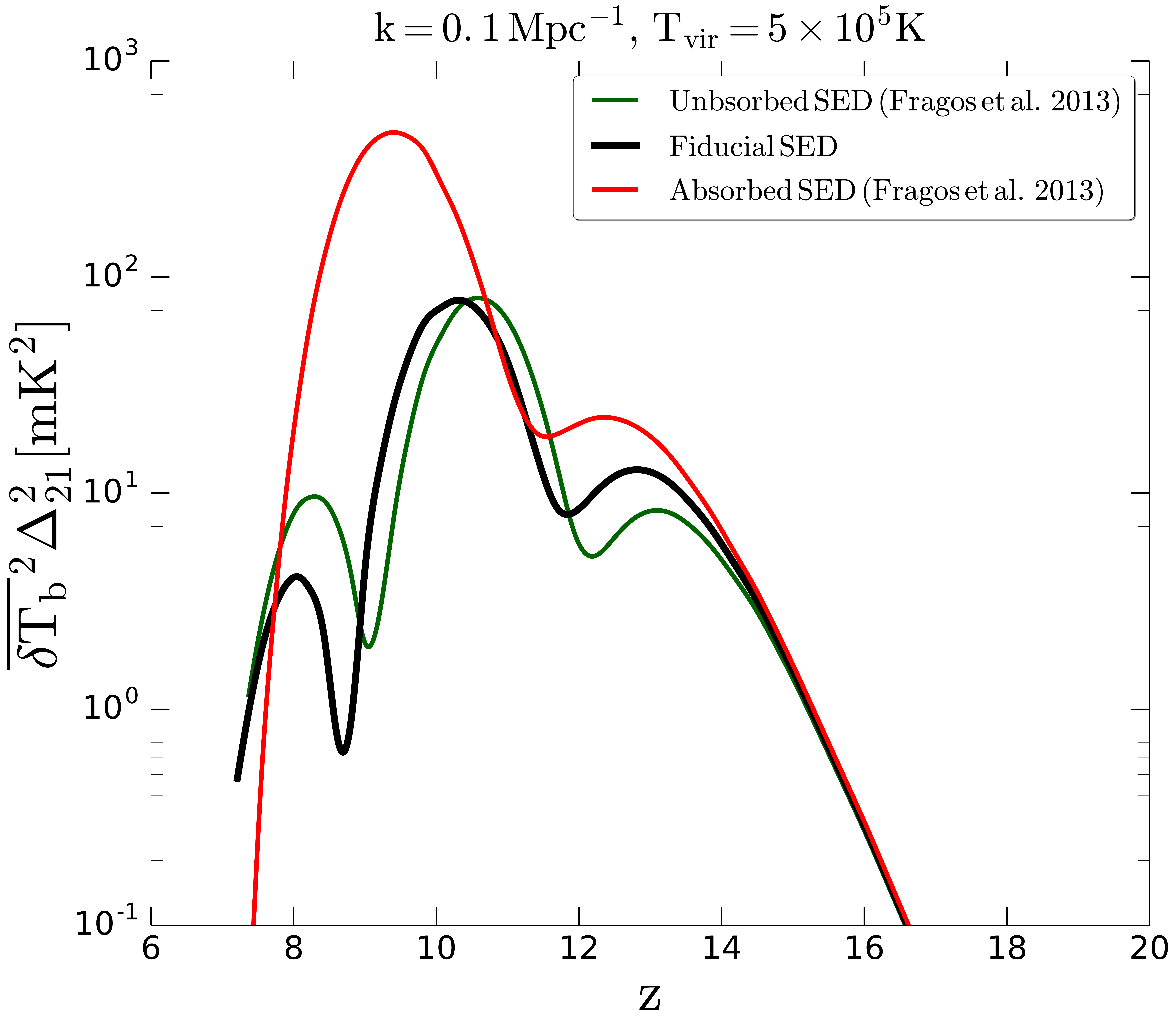}
 \includegraphics[width=0.45\textwidth]{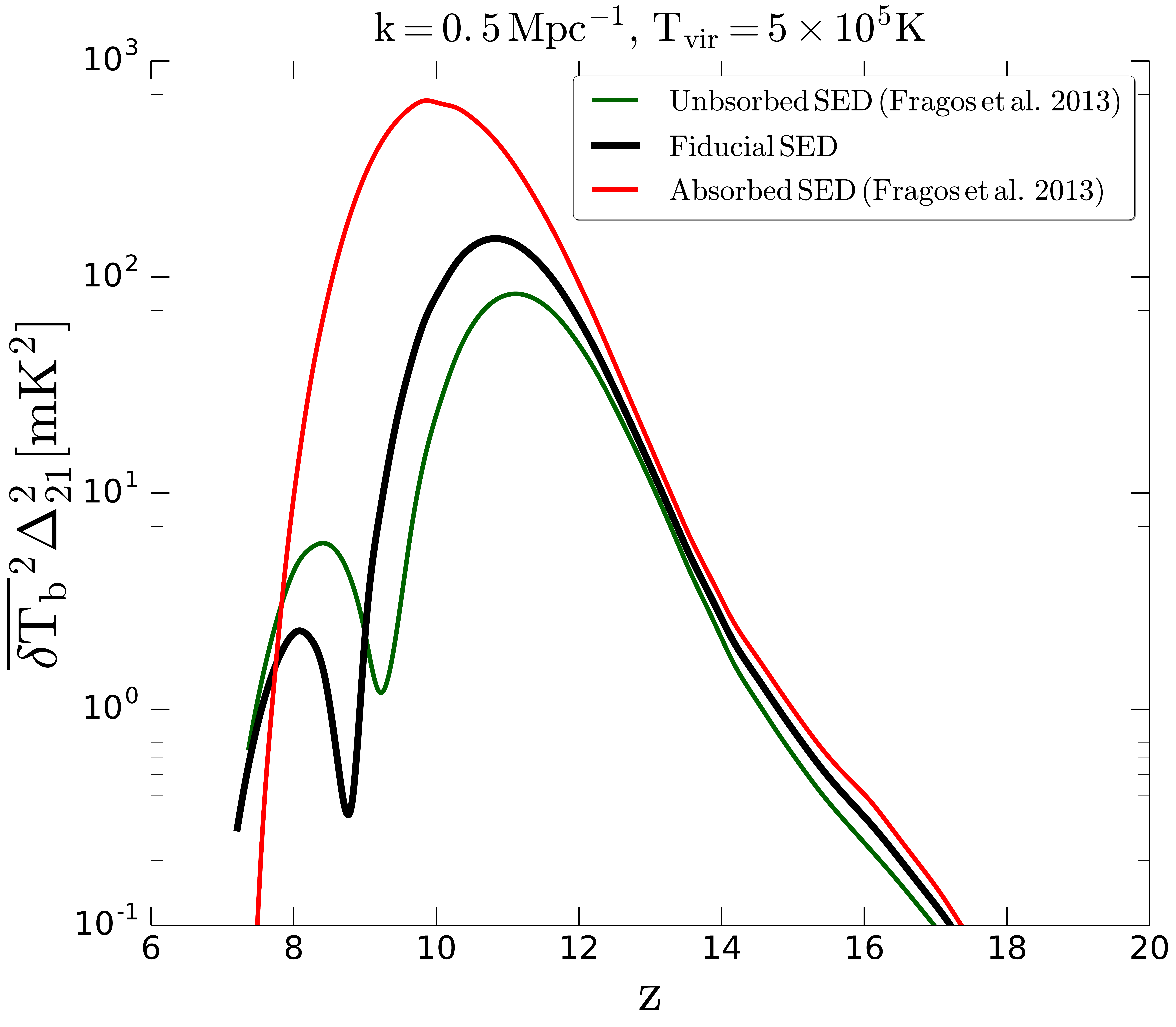}
}
\caption{
 Evolution of the spherically-averaged 21-cm power spectra, computed at $k=0.1$ Mpc$^{-1}$ ({\it left panels}) and $k=0.5$ Mpc$^{-1}$ ({\it right panels}). In the top row, we assume our fiducial model with $\Tvirmin=5\times10^4$ K, while in the bottom row we assume $\Tvirmin=5\times10^5$ K. In all panels, we present the evolution of the power assuming the three SEDs shown in Fig. \ref{fig:SED}. The blue curve in the top left panel corresponds to a simple toy-model SED: a power-law with energy index $\alpha = -1$, truncated below $E<0.5$ keV, whose amplitude is set by matching the soft-band luminosity to our fiducial SED.
\label{fig:PS}
}
\vspace{-0.5\baselineskip}
\end{figure*}

This is quantified in Fig. \ref{fig:PS}, which shows the redshift evolution of the 3D 21-cm power spectrum, corresponding to the three SEDs from Fig. \ref{fig:SED}. Focusing on results using the fiducial SED ({\it black curves}), we recover several previously-noted trends across the four panels: (i) the large-scale power exhibits a three-peaked evolution, driven by fluctuations in the WF coupling, gas temperature, and ionization ({\it left to right}); (ii) the middle peak, corresponding to the EoH, has the largest amplitude; (iii) rarer, more biased galaxies (denoted with a higher $\Tvirmin$) result in this evolution being delayed and then occurring more rapidly; (iv) if the EoH extends into the EoR (as is the case with a high $\Tvirmin$), the signal during the EoR is decreased owing to the $(1 - \Tcmb/\Ts)$ factor in eq. \ref{eq:delT} not saturating to unity until the end stages of the EoR.

The impact of the X-ray SED on these trends can be seen by comparing the different curves in each panel. Not including any absorption ({\it green curves}) results in a slightly earlier EoH compared to our fiducial model ({\it black curves}), due to the additional soft photons escaping the galaxy. However, this shift in the power spectrum is relatively minor, $\Delta z \sim 1$, given the modest X-ray opacities of our simulated galaxies.

In contrast, including the absorption in the \citet{Fragos12} model has a much more dramatic effect ({\it red curves}), as noted by \citet{FBV14}. This SED, calibrated to local HMXBs, results in most of the photons below $\lsim 2$ keV being absorbed in the host galaxy (c.f Fig. \ref{fig:SED}). The resulting dearth of soft photons results in a very uniform and inefficient heating of the IGM. This is evidenced by a factor of $\sim3$ reduction in the amplitude of the EoH peak in large-scale power for the $\Tvirmin=5\times10^4$ K model. For the more extreme $\Tvirmin=5\times10^5$ K model, the inefficient heating means that the EoH and EoR power spectrum peaks merge, as the first half of the EoR occurs while the IGM is still colder than the CMB and thus seen in absorption (cf. Fig. \ref{fig:temp}). The resulting ``cold reionization'' \citep{ME-WH14} is evidenced with a single, high-amplitude peak in large-scale 21-cm power, driven by the large contrast between the cold, neutral IGM (with a $\delT\sim$ -150 mK), and the ionized IGM (with a $\delT\sim$ 0 mK).

Finally, in the upper left panel, we also include the power spectrum evolution corresponding to a simplified SED: a power-law with energy index $\alpha = -1$, truncated below $E<0.5$ keV, whose amplitude is set by matching the soft-band luminosity to our fiducial SED. As expected from Fig. \ref{fig:SED}, a simple truncated power-law reproduces the results using our fiducial SED remarkably well (comparing the blue and black curves in the top left panel). A truncated power-law is the default X-ray SED in \cmfast, as it is computationally inexpensive to evaluate along the light-cone. Our results here support this commonly-used simplification.

\section{Conclusions}\label{sec:conc}

Upcoming 21-cm interferometers will allow us to study the first billion years of our Universe in unprecedented detail. The strongest cosmic signal is expected to come from the Epoch of Heating, which is expected to precede cosmic reionization. X-rays, likely from HMXBs, heat the IGM to temperatures above the CMB, driving the signal from absorption to emission.

In this work we improve on prior estimates of this process, quantifying the X-ray opacities in high-resolution simulations of the first galaxies. Those X-ray photons absorbed inside the ISM cannot escape into the IGM and thus contribute to the EoH. If this host galaxy absorption is strong, the heating of the IGM is inefficient and can dramatically impact the 21-cm signal.

We find that the typical X-ray opacity through our simulated galaxies can be approximated by a metal-free ISM with a typical column density of $N_{\rm HI}=10^{21.4}$ cm$^{-2}$. This absorption is modest compared to some previously-used values, based on local HMXBs.

Combining our simulated ISM opacities with public spectra of HMXBs, we compute the resulting 21-cm signal. Our results are consistent with ``standard scenarios'' in which the X-ray heating of the IGM is inhomogeneous, and occurs before the bulk of reionization. In our fiducial model, the large-scale ($k\sim0.1$ Mpc$^{-1}$) 21-cm power reaches a peak of $\approx 100$ mK$^2$ at $z\sim15$, well before the bulk of the EoR. Our results for the cosmic 21-cm signal can be reproduced assuming a simple power-law spectrum for the X-ray emission from HMXB, truncated at energies below 0.5 keV.


\section*{Acknowledgments}

We thank R. Salvaterra for helpful discussions. This project has received funding from the European Research Council (ERC) under the European Union's Horizon 2020 research and innovation programme (grant agreement No. 638809 -- AIDA -- PI: Mesinger). JHW acknowledges support from National Science Foundation (NSF) grants AST-1333360 and AST-1614333 and Hubble theory grants HST-AR-13895 and HST-AR-14326.

\bibliographystyle{mnras}
\bibliography{ms}

\end{document}